\documentclass[10pt,journal,compsoc]{IEEEtran}

\ifCLASSOPTIONcompsoc
  \usepackage[nocompress]{cite}
 \usepackage[caption=false,font=footnotesize,labelfont=sf,textfont=sf]{subfig}
\else
 \usepackage[caption=false,font=footnotesize]{subfig}
  \usepackage{cite}
\fi
\ifCLASSOPTIONcaptionsoff
 \usepackage[nomarkers]{endfloat}
\let\MYoriglatexcaption\caption
\renewcommand{\caption}[2][\relax]{\MYoriglatexcaption[#2]{#2}}
\fi
\usepackage{url}
\usepackage{fixltx2e}
\usepackage{enumitem}
\usepackage{array}
\usepackage{algorithmic}
\usepackage{amsmath}
\usepackage{graphicx}
\usepackage{tabularx}
\usepackage{multirow}
\usepackage{moreverb}
\usepackage{rotating}
\usepackage{tcolorbox}
\usepackage{dialogue}
\usepackage{float}
\usepackage{hyperref}
\usepackage{cleveref}
\usepackage{footnote}
\usepackage{mathrsfs}
\usepackage{booktabs}
\usepackage[htt]{hyphenat}
\usepackage{dblfloatfix}
\usepackage{wrapfig}
\usepackage{fancyvrb}
\usepackage{ragged2e}
\usepackage{multicol}
\usepackage{soul}
\usepackage{amssymb}

\usepackage{cuted}
\usepackage{capt-of}

\usetikzlibrary{positioning}

\newcounter{stepnode}
\counterwithin{stepnode}{figure} 
\renewcommand{\thestepnode}{\arabic{stepnode}} 
\newcommand{\setstep}[1]{\refstepcounter{stepnode}\label{#1}\thestepnode}

\newfloat{excerpt}{h}{lob}
\floatname{excerpt}{Excerpt}
\Crefname{excerpt}{Excerpt}{Excerpts}
\newcolumntype{Y}{>{\centering\arraybackslash}X}

\usepackage{mdframed}
\newmdenv[
  backgroundcolor=gray!10,
  linecolor=gray,
  linewidth=1pt,
  roundcorner=5pt,
  skipabove=10pt,
  skipbelow=10pt,
]{highlightbox}

\newcounter{finding}

\begin{document}

\title{Semantic Neighborhood Density and Eye Gaze Time in Human Programmer Attention}

\author{Robert Wallace, Emory Michaels, Yu Huang, Collin McMillan
\IEEEcompsocitemizethanks{
  \IEEEcompsocthanksitem Manuscript received ---- ---- -----. This work is supported in part by the NSF CCF-2100035 and CCF-2211428. Any opinions, findings, and conclusions expressed herein are the authors’ and do not necessarily reflect those of the sponsors.
  \IEEEcompsocthanksitem Authors Wallace, Michaels, and McMillan are with the Department
  of Computer Science and Engineering, University of Notre Dame. 
  E-mail: \{rwallac1, esmith36, cmc\}@nd.edu. 
  Huang is with  Department
  of Computer Science, Vanderbilt University. 
  E-mail: yu.huang@vanderbilt.edu
  \IEEEcompsocthanksitem This paper has supplementary downloadable multimedia material available
  at \url{https://github.com/apcl-research/attention-semantics}.
}}

\IEEEtitleabstractindextext{%
\begin{abstract}
This paper studies the relationship between human eye gaze time on words in source code and the Semantic Neighborhood Density (SND) of those words.  Human eye gaze time is a popular way to quantify human attention such as the importance of words people read and the cognitive effort people exert.  Meanwhile, SND is a measure of how similar a word is in meaning to other words in the same context.  SND has a long history in Psychology research where it has been connected to eye gaze time in various domains and helps explain human cognitive factors such as confusion and quality of reading comprehension.  But SND carries an unknown and potentially unique meaning in software engineering.  In this paper, we compute SND for tokens in source code that people viewed in two previous eye-tracking experiments, one in C and one in Java.  We conduct a model-free analysis for statistical relationships between SND and gaze time, and a model-based analysis for predictive power of SND to gaze time.  We found that words with high SND tend to have higher gaze time then low SND words, especially for words that are uncommon (i.e., have low frequency).  We also found SND and frequency to have a minor predictive power on gaze time, despite high levels of noise common in eye tracking data.

\end{abstract}

}

\maketitle

\section{Introduction}
\label{sec:intro}

This paper studies the relationship between human eye gaze time on words in source code and the Semantic Neighborhood Density (SND) of those words.  Human eye gaze time is a popular way to quantify factors related to human attention such as the importance of words people read and the cognitive effort people exert.  Quantified measures of human attention have long been a target of academic research as a study of the mind~\cite{mole2025encylopedia} and are increasing in relevance as they are used to manage user interface design, staffing, and, very recently, the design of machine attention systems in artificial neural networks~\cite{novak2024eye,baharum2024enhancing,cheng2012eye,zhang2024eyetrans,zhang2025enhancing, pourhosein2025unveiling,kiseleva2020study}.  In short, better understanding of human programmer attention leads to better software design.

Semantic Neighborhood Density is a measure of how similar a word is in meaning to other words in the same context~\cite{Buchanan_2001a}.  For example, in the context of the source code of a typical program, a word like {\small\texttt{array}} has high SND because it is similar in meaning to other common words like {\small\texttt{vector}}, {\small\texttt{list}}, and {\small\texttt{set}}.  But words like {\small\texttt{socket}} and {\small\texttt{sigmoid}} have low SND because they have very specific meanings with few synonyms in code.  SND has a long history in Psychology research where it is used to help explain human cognitive factors such as confusion, quality of reading comprehension, and concept recognition~\cite{Buchanan_2001a}\cite{Mirman_2008}. Yet it has been overlooked in a software engineering context despite program comprehension being a very active research area~\cite{maalej2014comprehension}\cite{schroter2017comprehending}.

SND carries an unknown and potentially unique meaning in software engineering.  Research on SND's effect on gaze time for \textit{concrete} words in a general, non-code reading context (i.e., words like ``hammer'' and ``nail'') has shown that dense semantic neighborhoods (high SND) slows reading as measured by eye gaze time because multiple semantically similar words are activated in the person's memory which the person must filter.  On the other hand, for \textit{abstract} words like ``justice'' and ``grace'' the effect is different: high SND abstract words seem to be easier to understand and have lower gaze time~\cite{Danguecan_2016}.  However, as Reilly and Desai~\cite{REILLY201746} point out, the effect on abstract words could be due to the emotional arousal these words cause.  Software engineering offers a unique environment where many words represent abstract concepts, but these concepts have can have concrete metaphors and are devoid of emotional weight.

In this paper, we study the effect of Semantic Neighborhood Density on words read by human programmers.  We obtained data for two eye tracking experiments: one in which programmers read Java source code and wrote documentation for that code, and a second in which programmers read C source code to find and explain memory bugs in that code.  The Java experiment involved 60 hours of eye tracking observations (10 programmers for 6 hours each) and the C experiment involved 31 hours (21 programmers for 1.5 hours each).  We adapted two methods of measuring SND for natural language to the Java and C programming languages.  Then we compared these measures to eye gaze time and controlled for the potentially confounding factors of word frequency and word predictability.

We found that high SND words are associated with higher values in metrics of eye gaze time, especially in the C dataset and when using a version of GPT2 we customize to compute SND.  The association is strongest among low frequency words.  In short, rarer words with higher SND tend to receive more visual attention than other words.


\section{Background and Related Work}

This section discusses key technologies and related work, such as studies of human attention, semantic neighborhood density, and human attention prediction using SND.

Figure \ref{tab:summarization} has key related work and background related items. We have divided them into five categories: Gaze-based, Natural Language, Code, Semantic Neighborhood Density, and Frequency metrics. Gaze-based refers to papers that used eye-tracking to observe the participant's behavior during their study (column \textit{G}). Papers in the Natural Language (column \textit{NL}) cover papers studying how participants read in their natural language rather than code. Papers in the Code category refer to reading software source code (column \textit{C}). Semantic Neighborhood Density (SND) and Frequency refer to papers that study SND or frequency metrics (columns \textit{SND} and \textit{F}). 

\begin{figure}[h]
\centering
\small
\vspace{-2mm}
\begin{tabular}{lllllll}
 & G & NL & C & SND & F & \\
Burgess (1998) \cite{Burgess1998} &  &  &  &  &  &  \\
Buchanan et al. (2001) \cite{Buchanan_2001a} &  & x &  & x & x &  \\
Mirman \& Magnusson (2008) \cite{Mirman_2008} &  & x &  & x &  &  \\
Shaoul \& Westbury (2010) \cite{Shaoul_2010a} &  & x &  & x &  &  \\
Al Farsi (2014) \cite{AlFarsi2014}                              & x & x &  & x & x &  \\
Rodeghero \textit{et al.} (2014) \cite{rodeghero2014improving}  & x &  & x &  &  &  \\
Pennington \textit{et al.} (2014)\cite{Pennington2014GloVeGV} &  & x &  &  &  &  \\
Al Madi \textit{et al.} (2021)\cite{al2021novice} & x &  & x &  & x & \\

Abid \textit{et al.} (2019)\cite{abid2019developer} & x &  & x \\ 
Sharafi \textit{et al.} (2020)\cite{Sharafi2020PracticalGuide} & x & x & x &  &  & \\
Wallace \textit{et al.} (2025)\cite{Wallace2025Programmer} & x &  & x &  &  & \\
Smith \textit{et al.} (2025)\cite{Smith_2025} & x &  & x &  &  & \\

\textless\textit{this paper}\textgreater & x &  & x & x & x & \\
\end{tabular}%
\caption{Overview of related work.  Column G means gaze-based. NL means the paper focused on natural language. C means the paper focused on code. SND means the paper studies semantic neighborhood density. F means the paper studies at frequency metrics. }
\vspace{-15px}
\label{tab:summarization}
\end{figure}

\subsection{Studies of Human Programmer Attention}
\label{sec:humanattn}

Human programmer attention refers to the allocation of a programmer's perceptual resources during software engineering tasks.  Usually research literature studies programmer attention to source code.  By far the most common type of attention studied, and the way we use in this paper, is specifically \textit{visual} attention that can be measured using eye tracking devices.  Eye tracking has been a staple of studies of programmer attention for decades, as chronicled in recent surveys~\cite{grabinger2024eye, Obaidellah2018, sharif2011use, Sharif2017Traceability} and for which empirically-validated guides on best practices exist~\cite{Sharafi2020PracticalGuide}.  

Two recent studies stand out as relevant because we use their data in this paper: one by Smith~\emph{et al.}~\cite{Smith_2025} and one by Wallace~\emph{et al.}~\cite{Wallace2025Programmer} to which we will refer as the ``Smith study'' and the ``Wallace study.''  The Smith study involved 21 programmers who read the source code of three different software projects in the C programming language.  The programmers read the C projects for around 1.5 hours to locate memory bugs.  As noted in Table~\ref{tab:dataset_comparison}
, the Wallace study involved 10 programmers who read the source code of five software projects in the Java programming language.  The programmers read the Java projects for around five hours to write programmer-level documentation for 40 methods in those projects.  The Smith study includes around 31 hours of eye tracking data and the Wallace study includes around 60 hours.  Both studies used a 120 Hertz Tobii Fusion Pro device and the iTrace toolkit~\cite{behler2023itracetoolkit} for eye tracking data collection.

\begin{figure}[t]
    \centering
{
\footnotesize
\setlength{\tabcolsep}{3pt} 
\renewcommand{\arraystretch}{0.0}
\setlength{\extrarowheight}{-8pt}
\begin{tabular}{c}

\includegraphics[width=6.0cm]{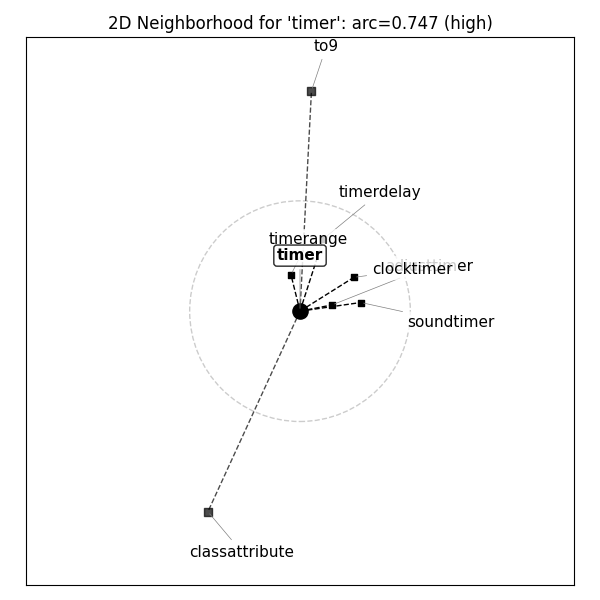} \\
\includegraphics[width=6.0cm]{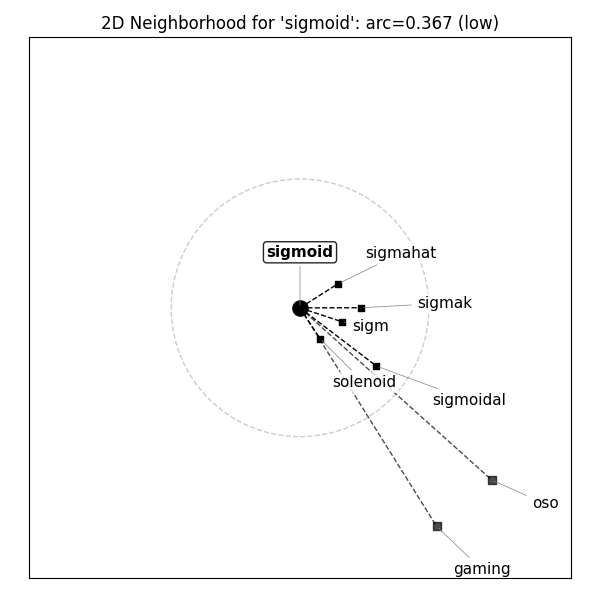} 

\end{tabular}
}
\vspace{-5px}
\caption{Example of words with high SND (upper) and low SND (lower) in source code from the Wallace study using GPT2's vector space.  The word ``timer'' lives in a more crowded semantic space than ``sigmoid'', which could have an effect on comprehension of those words.}
\label{fig:snd_example}
\vspace{-14px}
\end{figure}

\subsection{Semantic Neighborhood Density}
\label{sec:backsnd}

Semantic Neighborhood Density (SND) is a way to measure the similarity of words relative to other words in a corpus of documents~\cite{Buchanan_2001a}.  Consider all words in a corpus projected into an $n$-dimensional vector space.  Some words will be near many other words in this vector space.  Those words are defined as having more ``semantic neighbors.''  A measure of SND will be higher when a word's neighbors are closer to that word, and lower when neighbors are farther away.  There are many interpretations of high and low SND, but in general words that are used generically in many contexts will have high SND, whereas words with specific meanings used in specialized contexts will have low SND.

In code, high SND words might include ``process'', ``output'', and ``buffer.''  Low SND words might be ``keystore'', ``pagefault'', and ``websocket.''  Consider the example in Figure~\ref{fig:snd_example}. The neighbors of ``sigmoid'' are further away on average and quickly diverge in meaning, while the neighbors of ``timer''are closer and have similar meanings. We also include two words from beyond the semantic neighborhood defined in \ref{sec:snd} to illustrate semantically distant words.

The first measures of SND were published in the late 1990s~\cite{Burgess1998} and evolved over the following decade~\cite{Mirman_2008}.  In this paper we use the approach defined by Shaoul \textit{et al.} in 2010~\cite{Shaoul_2010a}, which has since been empirically validated in several domains~\cite{AlFarsi2014, Mirman_2008, winter2024iconicity} and become the de facto standard.  Please see Section~\ref{sec:snd} for our formal definition.

\begin{figure*}[!t]

\centering
\resizebox{\textwidth}{!}{
\begin{tikzpicture}[
    node distance=1cm and 1.2cm,
    >=stealth,
    num/.style={label={[font=\small, inner sep=1pt, anchor=north west]south east:\setstep{#1}}},
    block/.style={rectangle, draw, rounded corners, minimum width=2.4cm, minimum height=0.9cm, align=center, fill=white, font=\small},
    database/.style={rectangle, draw, shape border rotate=90, minimum width=1.4cm, minimum height=1.5cm, align=center, fill=white, font=\small},
    oval/.style={rectangle, draw, minimum width=2cm, minimum height=0.9cm, align=center, fill=white, font=\small},
    wideblock/.style={rectangle, draw, rounded corners, minimum width=4.2cm, minimum height=1.3cm, align=center, fill=white, font=\small}
]

    \node[database, num=step:repo] (repo) {Code\\Repository};
    \node[block, above=0.8cm of repo, num=step:lms] (lms) {Language\\Models};
    \node[oval, right=of lms, num=step:embs] (embs) {Word Embs.};
    
    \node[block, right=of repo, num=step:sndcalc] (sndcalc) {SND Calc};
    \node[block, below=0.6cm of sndcalc, num=step:freqcalc] (freqcalc) {Frequency Calc};

    \node[block, right=of sndcalc, num=step:sndword] (sndword) {SND/word};
    \node[block, right=of freqcalc, num=step:freqword] (freqword) {Freq/word};

    \node[block, right=of sndword, num=step:sndsplit] (sndsplit) {high/low\\split};
    \node[block, right=of freqword, num=step:freqsplit] (freqsplit) {high/low\\split};

    \node[database, below=2cm of repo, num=step:eyedata] (eyedata) {Eye\\Data};
    \node[block, right=of eyedata, num=step:eyecalc] (eyecalc) {eyetracking\\metrics calc};
    
    \node[wideblock, right=of eyecalc, num=step:metrics] (metrics) {
        single/first fixation, \\
        gaze duration, \\
        regression path duration
    };

    \node[block, right=1cm of freqsplit.south east, anchor=north west, num=step:modelfree] (modelfree) {Model-Free\\Analysis};
    \node[block, below=0.7cm of modelfree, num=step:modelbased] (modelbased) {Model-Based\\Analysis};

    \node[block, right=of modelfree] (resxy) {Results\\(Section \ref{sec:model-free-results})};
    \node[block, right=of modelbased] (resxz) {Results\\(Section \ref{sec:model-based-results})};

    \draw[->] (repo) -- (lms);
    \draw[->] (lms) -- (embs);
    \draw[->] (embs.south) -| (sndcalc.north);
    \draw[->] (repo.east) -- (sndcalc.west);
    \draw[->] (repo.east) -- ++(0.4,0) |- (freqcalc.west);
    \draw[->] (sndcalc) -- (sndword);
    \draw[->] (sndword) -- (sndsplit);
    \draw[->] (freqcalc) -- (freqword);
    \draw[->] (freqword) -- (freqsplit);
    \draw[->] (eyedata) -- (eyecalc);
    \draw[->] (eyecalc) -- (metrics);
    \draw[->] (sndsplit.east) -- ++(0.5,0) |- (modelfree.west);
    \draw[->] (sndsplit.east) -- ++(0.5,0) |- (modelbased.west);
    \draw[->] (freqsplit.east) -- ++(.7,0) |- ([yshift=2mm]modelfree.west);
    \draw[->] (freqsplit.east) -- ++(.7,0) |- ([yshift=2mm]modelbased.west);
    \draw[->] (metrics.east) -- ++(0.5,0) |- ([yshift=-2mm]modelfree.west);
    \draw[->] (metrics.east) -- ++(0.5,0) |- ([yshift=-2mm]modelbased.west);
    \draw[->] (modelfree) -- (resxy);
    \draw[->] (modelbased) -- (resxz);

\end{tikzpicture}
}
\captionof{figure}{Overview of our study showing components leading to model-free and model-based analyses.  The text of Section~\ref{sec:studydesign} is organized around and describes the items at the numbered labels in this figure.  Note we conduct separate analysis of each dataset in Section~\ref{sec:datasetpreprocessing}.}
\label{fig:pipeline}
\vspace{-2mm}
\end{figure*}
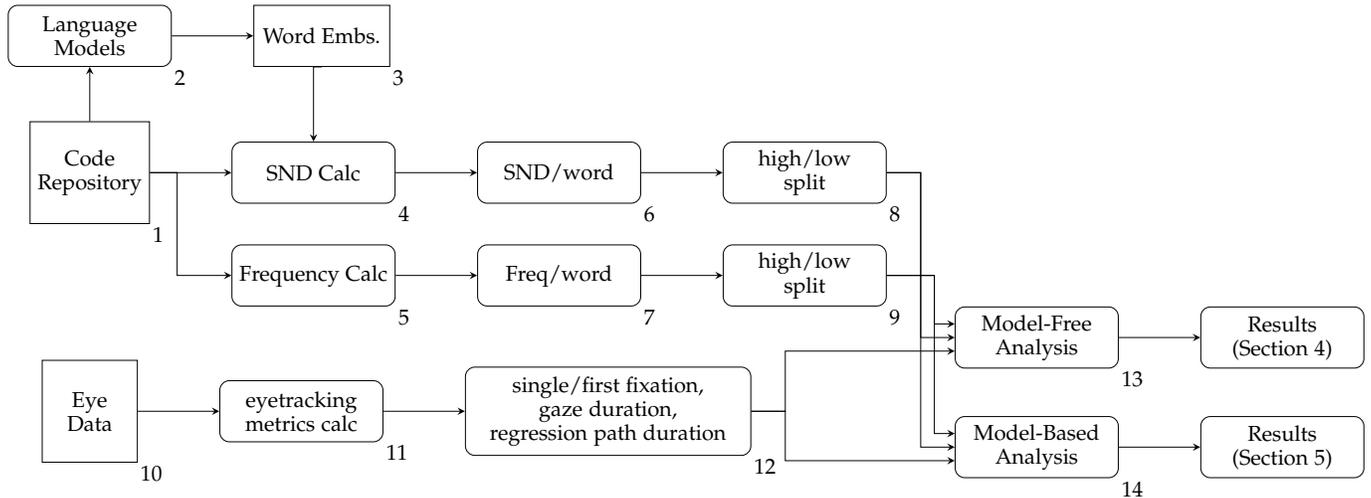

\subsection{Human Attention Prediction using SND}
One reason SND is considered useful is that it has been found to predict human attention, especially in conjunction with other metrics.  For example, Buchanan~\textit{et al.}~\cite{Buchanan_2001a} found that in reading English prose, eye gaze time is higher for words that have high SND if those words also have relatively low frequency.  Later, Danguecan and Buchanan~\cite{Danguecan_2016} found differences for concrete versus abstract words, also in English prose.  Concrete words with high SND were found to have longer eye gaze time, while abstract words with high SND had lower gaze time.  This finding was confirmed by Reilly and Desai~\cite{REILLY201746} who suggested emotional arousal of abstract words as a possible factor.  Other studies have linked SND to reading time in several domains such as second language learning~\cite{giraldez2020effect}, studies of aging~\cite{ayasse2020two, harel2021age}, and effects on sleep and memory~\cite{tamminen2013role}.





Especially relevant to this paper is dissertation work by Al Farsi~\cite{AlFarsi2014} that presented four experiments about SND interaction with lexical text metrics such as word frequency and orthographic neighborhood size to predict fixation duration during single-sentence reading. Experiment 1 showed that, controlling for factors such as frequency and predictability, high SND target words embedded in identical sentence frames generally produced shorter fixation durations than low SND words. Experiment 2 varied word frequency and found that high SND explained fixation duration more for high-frequency than low-frequency words, alongside effects of frequency and word length. Experiment~3 manipulated SND and word frequency while controlling other lexical variables and found limited evidence for SND effects. Experiment 4 showed that SND interacted with orthographic neighborhood size, with significant effects on first fixation, gaze duration and regression path duration, indicating that these text metrics jointly influence lexical identification during sentence reading.  

\vspace{-2mm}
\subsection{Novelty Statement}
This paper is novel because it studies eye gaze time with respect to SND and term frequency, in the domain of software engineering (Figure~\ref{tab:summarization}, final row).  Our studies' scope covers the C and Java languages and different SE tasks.

\section{Study Design}
\label{sec:studydesign}

Figure \ref{fig:pipeline} depicts our study design.  Our research objective is to study the association between SND and human eye gaze time on words in source code.  We also calculate and control for word frequency, in light of related work (namely, Al Farsi~\cite{AlFarsi2014}) indicating that high and low SND words can be further divided by high and low frequency of those words.  This section follows the flow we show in Figure~\ref{fig:pipeline}: datasets and preprocessing (area 1), language model word embeddings (areas 2 and 3), SND calculation (areas 4 through 9), eye tracking metrics (areas 10 through 12), and model-free and model-based analysis (areas 13 and 14).


\vspace{-2mm}
\subsection{Datasets and Preprocessing}
\label{sec:datasetpreprocessing}

\begin{table}[b]
\centering
\vspace{-3mm}
\caption{Comparison of C and Java Eye-Tracking Datasets}
\vspace{-2mm}
\label{tab:dataset_comparison}
\begin{tabular}{@{}lll@{}}
\textbf{} & Smith~\emph{et al.}~\cite{Smith_2025} & Wallace~\emph{et al.}~\cite{Wallace2025Programmer} \\ \midrule
Task Type & Bug Localization & Code Summarization \\
Language & C & Java \\
Participants & 21 & 10 \\
Data Recorded & 31 Hours & 60 Hours \\
Total Tasks & 8 Bug Reports & 40 Methods \\
Mean Fixation Count & 838 (per bug) & 263.5 (per method) \\
Regression Rate & 43.0\% -- 67.0\% & 49.8\% -- 56.6\% \\ \bottomrule
\end{tabular}
\end{table}

We use two datasets: the Smith study and Wallace study we discussed in Section~\ref{sec:humanattn}.  The Smith study includes C projects and the Wallace study includes Java projects that serve as code repositories (Figure~\ref{fig:pipeline}, area 1).  We preprocess each repository to extract words from the source code according to a lightweight process standard in SE literature~\cite{sun2014empirical}.  First we use srcML~\cite{collard2011lightweight} to extract all tokens according to the language specification of a token (i.e., C or Java).  For C, we use these source code tokens as the ``words'' in our study as-is.  For Java, we split identifier names by underscore and camel case, then convert all tokens to lower case, and these split/changed tokens are the ``words.''  We use this additional preprocessing step in Java based on recommendations from related work that highlight the effect naming conventions in each language have on the semantic content of words in those languages~\cite{butler2016analysing, herka2023identifier, scanniello2017fixing}.

\subsection{Language Model Word Embedding}
\label{sec:lm}

Areas 2 and 3 of Figure~\ref{fig:pipeline} show language models creating word embeddings for each of the words we preprocessed from word embeddings in the previous subsection.  We used two language models: GPT2 and CodeLLaMA.  The GPT2 model we used is from Su~\emph{et al.}~\cite{su2023language} who provide separate GPT2-like model versions for Java and C.  We use these versions because they are trained from scratch on vetted Java and C datasets which have known pretraining sets so we can retain tight control of experimental variables.  Also, these models are 350m parameters in size, which is a typical size for building word embeddings (e.g., BERT) and is a ``goldilocks'' size large enough to be competitive with very large commercial models on select program comprehension tasks, but small enough to run on consumer hardware without resolution loss (e.g., LoRA)~\cite{su2024distilled}.

We also use the 7 billion parameter version of CodeLLaMA, an open source model designed for software engineering tasks from Meta AI~\cite{roziere2023code}.  The advantage to CodeLLaMA is that it is much larger than our GPT2 version and has a larger pretraining set, but the disadvantage is that there is only one version for all languages.  So word embeddings from CodeLLaMA for C, for example, will be influenced by co-occurrence of those words in C, Java, Python, and even natural languages.
We use open source models so we can extract word embeddings ourselves without relying on commercial APIs which may make unknown changes to the embedding space.  The vector size of embeddings from GPT2 is 1024 and from CodeLLaMA is 4096.

\vspace{-1mm}
\subsection{Metrics Calculation}
\label{sec:snd}

This section describes how we calculate SND and frequency metrics, shown as areas 4 through 9 of our process in Figure~\ref{fig:pipeline}.  We calculate these metrics for each word in the vocabulary from each language.  We maintain a separate vocabulary for each dataset.  For each vocabulary, we calculate a global distance threshold around each word.  Then we collect all other words within that threshold in the vector space.  Then compute the similarity between those word vectors.  Then we average the similarities.  That average is the so-called Average Radius of Co-occurrence (ARC) by Shaoul~\emph{et al.}~\cite{Shaoul_2010a} that we use as our measure of SND:

Formally, let $\mathcal{V} \in \{\mathcal{V}_{C}, \mathcal{V}_{Java}\}$ denote a programming-language–specific vocabulary, and let $w \in \mathcal{V}$ denote a word. We map each word $w$ to a vector representation $\mathbf{x}_w \in \mathbb{R}^d$ using one of the language models described in Section~\ref{sec:lm}.

We then use the process defined by Shaoul and Wesbury to calculate the global distance threshold.  However, our vocabulary and embedding vector size makes it impractical to compute a threshold from all words.  So, similar to Al Farsi~\cite{AlFarsi2014}, we randomly sample 10k word pairs $(w_i,w_j)$.  Then like Shaoul and Wesbury we compute Euclidean distances $d_{ij} = \lVert \mathbf{x}_{w_i} - \mathbf{x}_{w_j} \rVert_2$ for word pairs in the vocabulary.
Where $\mu_d$ and $\sigma_d$ are the mean and standard deviation of these
distances, the global threshold is:

\vspace{-1mm}
\begin{equation}
\tau = \mu_d - 1.5\,\sigma_d
\end{equation}

We then define a ``semantic neighborhood'' around a word $w$ based on the threshold $\tau$ as:
\begin{equation}
\mathcal{N}(w) =
\left\{ y \in \mathcal{V} \setminus \{w\}
\;\middle|\;
\lVert \mathbf{x}_w - \mathbf{x}_y \rVert_2 \le \tau
\right\}
\end{equation}

Finally, we quantify SND using ARC, computed as the mean cosine similarity between $w$ and all words in its semantic neighborhood:
\begin{equation}
\mathrm{SND}(w) =
\mathrm{ARC}(w) =
\frac{1}{|\mathcal{N}(w)|}
\sum_{y \in \mathcal{N}(w)}
\frac{\mathbf{x}_w \cdot \mathbf{x}_y}
{\|\mathbf{x}_w\|\,\|\mathbf{x}_y\|}
\end{equation}

If a $w$ has no semantic neighbors, the token is categorized with no SND, which we treat as a low value.
Note that although ARC is sometimes defined in terms of average semantic distance, our definition instead uses average similarity within a thresholded neighborhood, yielding a measure that increases with semantic density.

We also calculate term frequency for each word.  Let $\mathcal{V} \in \{\mathcal{V}_{C}, \mathcal{V}_{Java}\}$ denote the
vocabulary, and let $f(w)$ denote the total number of occurrences of word
$w \in \mathcal{V}$ in the corpus. We define corpus-level term frequency as:
\begin{equation}
\mathrm{TF}(w) =
\frac{f(w)}{\sum_{w' \in \mathcal{V}} f(w')}
\end{equation}

We, like Al Farsi~\cite{AlFarsi2014}, use corpus-level TF because it provides an estimate of token prevalence in the repository, avoiding arbitrary document segmentation choices and aligning TF with our global embedding-based SND computation.  Corpus-level TF also has the advantage of being a proxy for a human reader's expected prior exposure to the term because a reader is more likely to be familiar with a word that occurs often over many software projects.




\vspace{-1mm}
\subsection{Eye Tracking Metrics}
\label{sec:eyemetrics}

This section describes the eye tracking metrics we calculate, corresponding to areas 10 through 12 in the overview Figure~\ref{fig:pipeline}.  The purpose of these eye tracking metrics is to quantify human visual attention levels.  To prepare, let $w$ denote a word and let $p$ denote a participant. Let $\mathcal{F}_{p}(w) = \{ f_1, f_2, \ldots \}$ denote the ordered sequence of fixations made by participant $p$ on word $w$, where each fixation $f_i$ has duration $\mathrm{dur}(f_i)$.  Then we compute these four metrics:

\vspace{1mm}

\textbf{Single Fixation Duration (SFD)} is the fixation duration on word $w$ in cases
where exactly one fixation is made:
\begin{equation}
\mathrm{SFD}_{p}(w) =
\begin{cases}
\mathrm{dur}(f_1), & |\mathcal{F}_{p}(w)| = 1, \\
\text{undefined}, & \text{otherwise}
\end{cases}
\end{equation}

\textbf{First Fixation Duration (FFD)} is the duration of the first fixation on word
$w$, regardless of whether additional fixations occur later:
\begin{equation}
\mathrm{FFD}_{p}(w) = \mathrm{dur}(f_1)
\end{equation}

\textbf{Gaze Duration (GD)} is the sum of all fixation durations on word $w$ prior to
the first saccade to another word:
\begin{equation}
\mathrm{GD}_{p}(w) =
\sum_{f_i \in \mathcal{F}^{\mathrm{first}}_{p}(w)} \mathrm{dur}(f_i)
\end{equation}

\textbf{Regression Path Duration (RPD)} is the total fixation time from the first
fixation on word $w$ until the participant’s gaze progresses beyond $w$ in the
reading order:
\begin{equation}
\mathrm{RPD}_{p}(w) =
\sum_{f_i \in \mathcal{F}^{\mathrm{reg}}_{p}(w)} \mathrm{dur}(f_i)
\end{equation}

All four are long-standing metrics in the eye tracking research community~\cite{clifton2007eye, rayner1998eye, rayner2004effect, warren2007investigating} and have been used in studies connecting eye gaze to SND in other domains~\cite{AlFarsi2014}.

\subsection{Model-Free Analysis}
\label{sec:modelfree}

This section covers our model-free analysis, depicted as area 13 in Figure~\ref{fig:pipeline}.  The purpose of this analysis is to quantify how SND and TF are associated with eye-tracking metrics independent of a learned prediction model.  First we partition the words into high and low groups using median splits.  Then we use a statistical test to compare the values of eye tracking metrics between groups.


\textbf{Vocabulary Partitioning}
Using above definitions of $V$, $SND(w)$, and $TF(w)$, we partition the vocabulary into high and low groups using median splits. Specifically, we define:
\begin{align}
V_{\text{HSND}} &= \{ w \in V \mid \mathrm{SND}(w) \ge \mathrm{median}(\mathrm{SND}) \}, \\
V_{\text{LSND}} &= V \setminus V_{\text{HSND}}, \\
V_{\text{HF}} &= \{ w \in V \mid \mathrm{TF}(w) \ge \mathrm{median}(\mathrm{TF}) \}, \\
V_{\text{LF}} &= V \setminus V_{\text{HF}}.
\end{align}


In addition to these marginal partitions, we define a joint category motivated by prior work showing interactions between SND and frequency~\cite{Buchanan_2001a}:
\begin{equation}
V_{\text{HSND,LF}} = V_{\text{HSND}} \cap V_{\text{LF}}.
\end{equation}
We group all remaining words as
\begin{equation}
V_{\text{Other}} = V \setminus V_{\text{HSND,LF}}.
\end{equation}

\textbf{Eye-Tracking Observations}
Next let a function $M \in \{\mathrm{SFD}, \mathrm{FFD}, \mathrm{GD}, \mathrm{RPD}\}$ denote one of the eye-tracking metrics defined in Section~\ref{sec:eyemetrics}, and let $M_p(w)$ denote the value of metric $M$ for participant $p$ on word $w$, when defined.

For a word group $G \subseteq V$ and participants $P$ from each study, we define the set of observations for metric $M$ as:
\begin{equation}
\mathcal{O}_M(G) = \{ M_p(w) \mid w \in G,\; p \in P,\; M_p(w) \neq \varnothing \}
\end{equation}

\textbf{Statistical Comparisons}
We perform pairwise comparisons of eye-tracking metrics between the following groups:
\begin{enumerate}
    \item $V_{\text{HSND}}$ vs.\ $V_{\text{LSND}}$,
    \item $V_{\text{HF}}$ vs.\ $V_{\text{LF}}$,
    \item $V_{\text{HSND,LF}}$ vs.\ $V_{\text{Other}}$.
\end{enumerate}

For each comparison, let $\mathcal{O}_M(G_1)$ and $\mathcal{O}_M(G_2)$ denote the observation sets for the two word groups being compared. We test for differences in mean gaze time using a permutation test of means applied to these sets. To reduce sensitivity to extreme values commonly observed in eye-tracking data, we winsorize both samples at fixed 5\% percentile cutoffs prior to testing.

Permutation testing is appropriate here because gaze-duration measures are typically non-Gaussian, heavy-tailed, and heteroskedastic, and may exhibit unequal sample sizes across conditions~\cite{rayner1998eye, staub2010eye}. Unlike parametric tests, permutation tests make no distributional assumptions beyond exchangeability under the null hypothesis and remain valid under skewed distributions and unequal variances~\cite{good2005permutation, ernst2004permutation}. This property is especially important in our setting, where fixation-based metrics could differ in scale and variance across tasks, participants, and word categories.

\textbf{Reporting} We report the $p$-value from the permutation test, the $p$-value corrected for false discovery rate using the Benjamini--Hochberg procedure, and Hedges' $g$ as a standardized effect size to quantify the magnitude of differences between groups while correcting for unequal sample sizes.

\subsection{Model-Based Analysis}

This section covers our model-based analysis, depicted as area 14 in Figure~\ref{fig:pipeline}.
Unlike the model-free analysis, which characterizes group-level differences, this model-based analysis assesses the extent to which these differences support predictive discrimination between word groups.



\textbf{Prediction Task Definition}
We use the same partitioning procedure as in the model-free analysis, but apply it separately for each eye-tracking metric. Let $\mu \in \{\mathrm{SFD}, \mathrm{FFD}, \mathrm{GD}, \mathrm{RPD}\}$ denote an eye-tracking measure. For each $\mu$, we partition the vocabulary into:
\begin{align}
V_{\text{H$\mu$}}^{m} &= \{ w \in V \mid \mathrm{SND}(w) \ge \mathrm{median}(\mathrm{SND}) \} \\
V_{\text{L$\mu$}}^{m} &= V \setminus V_{\text{H$\mu$}}^{m} \\
V_{\text{H$\mu$}}^{q} &= \{ w \in V \mid \mathrm{SND}(w) \ge Q_{0.75}(\mathrm{SND}) \} \\
V_{\text{L$\mu$}}^{q} &= \{ w \in V \mid \mathrm{SND}(w) \le Q_{0.25}(\mathrm{SND}) \}
\end{align}

For example, $V_{\text{HFFD}}^{q}$ means the set of words in the top quartile according to the eye tracking metric First Fixation Duration.  Note we create high and low split groups based on the median and upper/lower quartiles for each eye tracking metric.  The median split groups are larger, but samples near the median may be noisy in eye tracking data (e.g., a sample in the 45th percentile may be nearly indistinguishable in terms of SND from a sample in the 55th percentile).  Therefore we study upper/lower quartiles to minimize this potential noise around the median.

To create the comparison groups, we let $G_1^{\mu}$ and $G_2^{\mu}$ denote two word groups for measure $\mu$. We define a binary classification task in which the goal is to predict a word’s group membership for $\mu$, as follows:
\[
y_{\mu}(w) =
\begin{cases}
1 & \text{if } w \in G_1^{\mu}, \\
0 & \text{if } w \in G_2^{\mu}.
\end{cases}
\]
Each observation $\mu_p(w) \in \mathcal{O}_{\mu}(G_1^{\mu} \cup G_2^{\mu})$ serves as an input instance labeled by $y_{\mu}(w)$.

\textbf{Statistical Model}
For each eye-tracking metric $\mu$, we fit a generalized linear model (GLM) with a binomial response and logit link function. The model estimates the probability that a given observation belongs to group $G_1$ as a function of the corresponding gaze metric value:
\[
\Pr(y = 1 \mid \mu_p(w)) = \sigma(\beta_0 + \beta_1 \mu_p(w)),
\]
where $\sigma(\cdot)$ denotes the logistic function and $\beta_0, \beta_1$ are model parameters estimated via maximum likelihood.

GLM binomial models are used in eye-tracking and psycholinguistic research to relate continuous gaze-based measures to categorical experimental conditions because they provide a principled probabilistic framework, require minimal distributional assumptions on predictors, and yield interpretable effect estimates \cite{krajewski2010rh, jaeger2008categorical, fox2015applied}.

\textbf{Evaluation}
We train models in a leave-one-out (LOO) cross validation configuration where one word is left out one each round.  We use LOO because eye-tracking data are sparse and unevenly distributed, where LOO is the recommended~\cite{arlot2010survey}.  We create and evaluate separate models for each eye-tracking metric, word-group comparison, embedding, and dataset.  We report standard classification metrics including accuracy, precision, recall, F$_1$ score, and area under the ROC curve (AUC).

\begin{table*}[]
\centering
\caption{Model-free (statistical) comparison of high and low SND and TF groups for data from the Wallace study (Java).}
\vspace{-2mm}
\begin{tabular}{llrrrrrrr}
\hline
\multicolumn{1}{|l|}{Metric} & \multicolumn{1}{l|}{Comparison} & \multicolumn{1}{l|}{$n_1$} & \multicolumn{1}{l|}{$n_2$} & \multicolumn{1}{l|}{$\mu_1$} & \multicolumn{1}{l|}{$\mu_2$} & \multicolumn{1}{l|}{Hedges' $g$} & \multicolumn{1}{l|}{$p$-value} & \multicolumn{1}{l|}{$p$ (fdr)} \\ \hline
\multicolumn{9}{|l|}{GPT2-Java} \\ \hline
sfd & $V_{\text{HSND}}$ vs.\ $V_{\text{LSND}}$ & 656 & 671 & 408.827 & 400.926 & 0.053 & 0.172 & 0.172 \\
ffd & $V_{\text{HSND}}$ vs.\ $V_{\text{LSND}}$ & 657 & 673 & 413.583 & 403.079 & 0.071 & 0.101 & 0.172 \\
gd & $V_{\text{HSND}}$ vs.\ $V_{\text{LSND}}$ & 657 & 673 & 414.480 & 403.929 & 0.071 & 0.099 & 0.172 \\
rpd & $V_{\text{HSND}}$ vs.\ $V_{\text{LSND}}$ & 657 & 673 & 14822.765 & 13407.366 & 0.058 & 0.144 & 0.172 \\
sfd & $V_{\text{HF}}$ vs.\ $V_{\text{LF}}$ & 664 & 663 & 401.066 & 411.090 & 0.065 & 0.118 & 0.445 \\
ffd & $V_{\text{HF}}$ vs.\ $V_{\text{LF}}$ & 665 & 665 & 406.350 & 411.558 & 0.034 & 0.264 & 0.445 \\
gd & $V_{\text{HF}}$ vs.\ $V_{\text{LF}}$ & 665 & 665 & 408.927 & 412.128 & 0.021 & 0.354 & 0.445 \\
rpd & $V_{\text{HF}}$ vs.\ $V_{\text{LF}}$ & 665 & 665 & 13820.154 & 14002.757 & 0.008 & 0.445 & 0.445 \\
sfd & $V_{\text{HSND,LF}}$ vs.\ $V_{\text{Other}}$ & 191 & 1136 & 428.820 & 402.628 & 0.170 & \textbf{0.015} & \textbf{0.028} \\
ffd & $V_{\text{HSND,LF}}$ vs.\ $V_{\text{Other}}$ & 191 & 1139 & 431.217 & 406.199 & 0.165 & \textbf{0.017} & \textbf{0.028} \\
gd & $V_{\text{HSND,LF}}$ vs.\ $V_{\text{Other}}$ & 191 & 1139 & 431.832 & 407.184 & 0.162 & \textbf{0.021} & \textbf{0.028} \\
rpd & $V_{\text{HSND,LF}}$ vs.\ $V_{\text{Other}}$ & 191 & 1139 & 15480.334 & 13125.045 & 0.106 & 0.092 & 0.092 \vspace{2px}\\ \hline
\multicolumn{9}{|l|}{CodeLLaMA} \\ \hline
sfd & $V_{\text{HSND}}$ vs.\ $V_{\text{LSND}}$ & 648 & 679 & 409.967 & 400.795 & 0.061 & 0.132 & 0.132 \\
ffd & $V_{\text{HSND}}$ vs.\ $V_{\text{LSND}}$ & 649 & 681 & 413.401 & 403.900 & 0.064 & 0.120 & 0.132 \\
gd & $V_{\text{HSND}}$ vs.\ $V_{\text{LSND}}$ & 649 & 681 & 414.587 & 405.082 & 0.064 & 0.127 & 0.132 \\
rpd & $V_{\text{HSND}}$ vs.\ $V_{\text{LSND}}$ & 649 & 681 & 15081.317 & 13072.128 & 0.083 & 0.063 & 0.132 \\
sfd & $V_{\text{HF}}$ vs.\ $V_{\text{LF}}$ & 664 & 663 & 401.066 & 411.090 & 0.065 & 0.118 & 0.443 \\
ffd & $V_{\text{HF}}$ vs.\ $V_{\text{LF}}$ & 665 & 665 & 406.350 & 411.558 & 0.034 & 0.263 & 0.443 \\
gd & $V_{\text{HF}}$ vs.\ $V_{\text{LF}}$ & 665 & 665 & 408.927 & 412.128 & 0.021 & 0.353 & 0.443 \\
rpd & $V_{\text{HF}}$ vs.\ $V_{\text{LF}}$ & 665 & 665 & 13820.153 & 14002.757 & 0.007 & 0.442 & 0.442 \\
sfd & $V_{\text{HSND,LF}}$ vs.\ $V_{\text{Other}}$ & 272 & 1055 & 434.853 & 399.766 & 0.224 & \textbf{0.001} & \textbf{0.003} \\
ffd & $V_{\text{HSND,LF}}$ vs.\ $V_{\text{Other}}$ & 273 & 1057 & 434.764 & 404.304 & 0.196 & \textbf{0.002} & \textbf{0.004} \\
gd & $V_{\text{HSND,LF}}$ vs.\ $V_{\text{Other}}$ & 273 & 1057 & 434.515 & 405.572 & 0.186 & \textbf{0.003} & \textbf{0.004} \\

rpd & $V_{\text{HSND,LF}}$ vs.\ $V_{\text{Other}}$ & 273 & 1057 & 16564.472 & 13386.902 & 0.128 & \textbf{0.035} & \textbf{0.035} \\ \hline
\end{tabular}\vspace{4px}

\vspace{-4mm}
\label{tab:context-free-java}
\end{table*}

\begin{table*}[]
\centering
\caption{Model-free (statistical) comparison of high and low SND and TF groups for data from the Smith study (C).}
\vspace{-2mm}
\begin{tabular}{llllrrrrr}
\hline
\multicolumn{1}{|l|}{metric} & \multicolumn{1}{l|}{comparison} & \multicolumn{1}{l|}{$n_1$} & \multicolumn{1}{l|}{$n_2$} & \multicolumn{1}{l|}{$\mu_1$} & \multicolumn{1}{l|}{$\mu_2$} & \multicolumn{1}{l|}{Hedges' $g$} & \multicolumn{1}{l|}{$p$-value} & \multicolumn{1}{l|}{$p$ (fdr)} \\ \hline
\multicolumn{9}{|l|}{GPT2-C} \\ \hline
sfd &$V_{\text{HSND}}$ vs.\ $V_{\text{LSND}}$ &364 &1409 &343.824 &318.452 &0.157 &\textbf{0.005} &\textbf{0.009} \\
ffd &$V_{\text{HSND}}$ vs.\ $V_{\text{LSND}}$ &371 &1425 &344.593 &321.104 &0.146 &\textbf{0.007} &\textbf{0.009} \\
gd &$V_{\text{HSND}}$ vs.\ $V_{\text{LSND}}$ &371 &1425 &346.300 &322.160 &0.150 &\textbf{0.006} &\textbf{0.009} \\
rpd &$V_{\text{HSND}}$ vs.\ $V_{\text{LSND}}$ &371 &1425 &2384.950 &2832.592 &0.108 &0.972 &0.972 \\
sfd &$V_{\text{HF}}$ vs.\ $V_{\text{LF}}$ &888 &885 &314.332 &333.633 &0.118 &\textbf{0.006} &\textbf{0.014} \\
ffd &$V_{\text{HF}}$ vs.\ $V_{\text{LF}}$ &898 &898 &317.566 &335.198 &0.108 &\textbf{0.010} &\textbf{0.014} \\
gd &$V_{\text{HF}}$ vs.\ $V_{\text{LF}}$ &898 &898 &318.799 &336.815 &0.110 &\textbf{0.010} &\textbf{0.014} \\
rpd &$V_{\text{HF}}$ vs.\ $V_{\text{LF}}$ &898 &898 &2590.711 &3023.068 &0.099 &\textbf{0.019} &\textbf{0.019} \\
sfd &$V_{\text{HSND,LF}}$ vs.\ $V_{\text{Other}}$ &263 &1510 &344.889 &319.701 &0.156 &\textbf{0.011} &\textbf{0.017} \\
ffd &$V_{\text{HSND,LF}}$ vs.\ $V_{\text{Other}}$ &269 &1527 &346.521 &322.111 &0.152 &\textbf{0.013} &\textbf{0.017} \\
gd &$V_{\text{HSND,LF}}$ vs.\ $V_{\text{Other}}$ &269 &1527 &348.546 &323.448 &0.155 &\textbf{0.010} &\textbf{0.017} \\
rpd &$V_{\text{HSND,LF}}$ vs.\ $V_{\text{Other}}$ &269 &1527 &2653.963 &2770.156 &0.028 &0.656 &0.656 \\ \hline
\multicolumn{9}{|l|}{CodeLLaMA} \\ \hline
sfd &$V_{\text{HSND}}$ vs.\ $V_{\text{LSND}}$ &759 &1014 &320.303 &325.906 &0.035 &0.768 &0.768 \\
ffd &$V_{\text{HSND}}$ vs.\ $V_{\text{LSND}}$ &768 &1028 &323.145 &327.755 &0.029 &0.723 &0.768 \\
gd &$V_{\text{HSND}}$ vs.\ $V_{\text{LSND}}$ &768 &1028 &324.144 &328.973 &0.030 &0.729 &0.768 \\
rpd &$V_{\text{HSND}}$ vs.\ $V_{\text{LSND}}$ &768 &1028 &2789.472 &2736.040 &0.013 &0.396 &0.768 \\
sfd &$V_{\text{HF}}$ vs.\ $V_{\text{LF}}$ &888 &885 &314.332 &333.633 &0.118 &\textbf{0.006} &\textbf{0.016} \\
ffd &$V_{\text{HF}}$ vs.\ $V_{\text{LF}}$ &898 &898 &317.566 &335.198 &0.108 &\textbf{0.012} &\textbf{0.016} \\
gd &$V_{\text{HF}}$ vs.\ $V_{\text{LF}}$ &898 &898 &318.799 &336.815 &0.110 &\textbf{0.011} &\textbf{0.016} \\
rpd &$V_{\text{HF}}$ vs.\ $V_{\text{LF}}$ &898 &898 &2590.711 &3023.068 &0.099 &\textbf{0.019} &\textbf{0.019} \\
sfd &$V_{\text{HSND,LF}}$ vs.\ $V_{\text{Other}}$ &303 &1470 &342.108 &320.424 &0.133 &\textbf{0.020} &\textbf{0.028} \\
ffd &$V_{\text{HSND,LF}}$ vs.\ $V_{\text{Other}}$ &308 &1488 &343.829 &322.277 &0.134 &\textbf{0.018} &\textbf{0.028} \\
gd &$V_{\text{HSND,LF}}$ vs.\ $V_{\text{Other}}$ &308 &1488 &344.669 &324.028 &0.127 &\textbf{0.021} &\textbf{0.028} \\
rpd &$V_{\text{HSND,LF}}$ vs.\ $V_{\text{Other}}$ &308 &1488 &2803.999 &2748.684 &0.013 &0.412 &0.412 \\ \hline
\end{tabular}\vspace{4px}

\label{tab:context-free-c}
\vspace{-4.7mm}
\end{table*}

\section{Model-free Results}
\label{sec:model-free-results}

\begin{table*}[b]
\vspace{-2mm}
\centering
\caption{Model-based analysis showing differentiability based on medians ($V_{\text{H$\mu$}}^{m}$ vs. $V_{\text{L$\mu$}}^{m}$) for the Wallace study (Java).}
\vspace{-2mm}
\begin{tabular}{llllllll}
\hline
\multicolumn{1}{|l|}{Category} & \multicolumn{1}{l|}{Model} & \multicolumn{1}{l|}{Accuracy} & \multicolumn{1}{l|}{Precision} & \multicolumn{1}{l|}{Recall} & \multicolumn{1}{l|}{F1 Score} & \multicolumn{1}{l|}{ROC AUC} & \multicolumn{1}{l|}{Confusion Matrix} \\ \hline
\multicolumn{8}{|l|}{GPT2-Java} \\ \hline
sfd & GLM & 52.82\% & 0.5482 & 0.3283 & 0.4106 & 0.5072 & TN=239, FP=89, FN=221, TP=108 \\
ffd & GLM & 49.77\% & 0.5 & 0.284 & 0.3622 & 0.488 & TN=234, FP=94, FN=237, TP=94 \\
gd & GLM & 51.59\% & 0.5353 & 0.2749 & 0.3633 & 0.4938 & TN=249, FP=79, FN=240, TP=91 \\
rpd & GLM & 61.91\% & 0.6426 & 0.5394 & 0.5865 & 0.6749 & TN=230, FP=99, FN=152, TP=178  \\ \hline 
\multicolumn{8}{|l|}{CodeLLaMA} \\ \hline
sfd & GLM & 48.92\% & 0.4878 & 0.3058 & 0.3759 & 0.5104 & TN=218, FP=105, FN=227, TP=100 \\
ffd & GLM & 48.69\% & 0.4856 & 0.3079 & 0.3769 & 0.5064 & TN=216, FP=107, FN=227, TP=101 \\
gd & GLM & 50.08\% & 0.507 & 0.3323 & 0.4015 & 0.5269 & TN=217, FP=106, FN=219, TP=109 \\
rpd & GLM & 60.86\% & 0.6577 & 0.4506 & 0.5348 & 0.6595 & TN=249, FP=76,\hspace{4px} FN=178, TP=146 \\ \hline

\end{tabular}\vspace{4px}
\label{tab:model-based-j}
\end{table*}

\begin{table*}[b]
\vspace{-3mm}
\centering
\caption{Model-based analysis showing differentiability based on medians ($V_{\text{H$\mu$}}^{m}$ vs. $V_{\text{L$\mu$}}^{m}$) for the Smith study (C).}
\vspace{-2mm}
\begin{tabular}{llllllll}
\hline
\multicolumn{1}{|l|}{Category} & \multicolumn{1}{l|}{Model} & \multicolumn{1}{l|}{Accuracy} & \multicolumn{1}{l|}{Precision} & \multicolumn{1}{l|}{Recall} & \multicolumn{1}{l|}{F1 Score} & \multicolumn{1}{l|}{ROC AUC} & \multicolumn{1}{l|}{Confusion Matrix} \\ \hline
\multicolumn{7}{|l|}{GPT2-C} & \multicolumn{1}{l|}{} \\ \hline
sfd & GLM & 57.07\% & 0.5761 & 0.8683 & 0.6926 & 0.5319 & TN=32,\hspace{4px} FP=131, FN=27, TP=178 \\
ffd & GLM & 56.28\% & 0.5691 & 0.8564 & 0.6838 & 0.526 & TN=33, FP=131, FN=29, TP=173 \\
gd & GLM & 56.10\% & 0.5686 & 0.8529 & 0.6824 & 0.5329 & TN=33, FP=132, FN=30, TP=174 \\
rpd & GLM & 60.73\% & 0.6132 & 0.8627 & 0.7169 & 0.5784 & TN=39, FP=111, FN=28, TP=176 \\ \hline
\multicolumn{8}{|l|}{CodeLLaMA} \\ \hline
sfd & GLM & 53.17\% & 0.5395 & 0.6181 & 0.5761 & 0.5552 & TN=165, FP=210, FN=152, TP=246 \\
ffd & GLM & 52.43\% & 0.5309 & 0.6418 & 0.5811 & 0.5357 & TN=152, FP=228, FN=144, TP=258 \\
gd & GLM & 52.49\% & 0.5317 & 0.6452 & 0.583 & 0.5395 & TN=151, FP=229, FN=143, TP=260 \\
rpd & GLM & 52.16\% & 0.5321 & 0.5116 & 0.5216 & 0.5694 & TN=199, FP=175, FN=190, TP=199 \\ \hline 
\end{tabular}\vspace{4px}
\label{tab:model-based-c}
\end{table*}

Tables~\ref{tab:context-free-java} and~\ref{tab:context-free-c} summarize the results of our model-free analysis for the Java (Wallace study) and C (Smith study) datasets, respectively. Across both languages, we observe systematic but nuanced relationships between Semantic Neighborhood Density (SND), term frequency (TF), and eye-tracking measures of programmer attention. The observed patterns depend on the specific eye gaze metric and interactions between SND and TF.  We also note associations of SND from GPT2 seem stronger in general than from CodeLLaMA, but are comparable overall.


\textbf{SND Effects} For the Java dataset, marginal comparisons between high- and low-SND words ($V_{\text{HSND}}$ vs.\ $V_{\text{LSND}}$) reveal small and generally non-significant differences across all four eye-tracking metrics, regardless of whether SND is computed using GPT-2 or CodeLLaMA embeddings. Mean fixation-based measures (SFD, FFD, and GD) are consistently higher for high-SND words.  However, effect sizes are small (Hedges' $g \approx 0.05$--$0.08$) and none of the comparisons remain significant after false-discovery-rate correction. These results do not support a conclusion that SND alone has a strong association with gaze behavior during Java code summarization tasks.

In contrast, the C dataset shows clearer effects of SND when SND is derived from GPT-2 embeddings. High-SND words are associated with significantly longer single fixation duration, first fixation duration, and gaze duration compared to low-SND words, with small-to-moderate effect sizes ($g \approx 0.14$--$0.16$) that remain significant after FDR correction. No corresponding effect is observed for regression path duration. When using CodeLLaMA embeddings, these SND-related effects are lower and do not reach statistical significance. This pattern suggests that SND effects are pronounced in the C bug localization task.  Also, these results argue in favor of the GPT2-based word embeddings, at least when computing SND using ARC.

\textbf{TF Effects}  In the Java dataset, high-frequency words tend to receive slightly shorter fixation durations than low-frequency words, but these differences do not reach statistical significance. In the C dataset, high-frequency words are associated with significantly shorter fixation-based measures (SFD, FFD, and GD) as well as shorter regression path durations, with effects surviving FDR correction under both embedding models. These findings are consistent with prior work showing that higher-frequency words generally require less processing effort~\cite{kliegl2004length, rayner1998eye, starke2009searching}.

\textbf{Interaction of SND and TF}
The most robust effects across both datasets arise when considering the joint category of high-SND, low-frequency words ($V_{\text{HSND,LF}}$) compared to all other words. In the Java dataset, this group exhibits significantly longer fixation-based measures across SFD, FFD, and GD for both GPT-2 and CodeLLaMA embeddings, with effect sizes in the small-to-moderate range ($g \approx 0.16$--$0.22$). Regression path duration follows the same directional trend, though statistical significance is weaker.

Similarly, in the C dataset, words in the high-SND, low-frequency category elicit longer fixation durations than other words across all fixation-based metrics, with statistically significant differences observed for both embedding models. As in the Java dataset, regression path duration does not show reliable differences between groups, suggesting that SND--TF interactions primarily influence early and local stages of lexical processing (measured by FFD and SND) rather than rereading behavior (measured by RPD).

\textbf{Summary}
Our strongest finding is that in source code, words with high SND and low TF tend to elicit more human visual attention than other words.  This finding is consistent with literature in generalized reading~\cite{AlFarsi2014}, about which we will expand in Section~\ref{sec:conclusion}.  Statistically speaking, this finding is consistently significant over both programming languages and embedding spaces, for SFD, FFD, and GD.  The connection to RPD is less clear, suggesting differences between eye fixation and regression behavior.  SND and TF alone have less effect, with higher gaze time on high SND words and higher gaze time on low TF words observed in most conditions, though these results are not consistently significant.  We also note differences in word embedding source (GPT2 or CodeLLaMA), with GPT2-based embeddings generally superior, though more so in C than Java.

\begin{table*}[]
\centering
\caption{Model-based analysis showing differentiability based on upper and lower quartiles ($V_{\text{H$\mu$}}^{q}$ vs. $V_{\text{L$\mu$}}^{q}$) for the Wallace study (Java).}
\vspace{-2mm}
\begin{tabular}{llllllll}
\hline
\multicolumn{1}{|l|}{Category} & \multicolumn{1}{l|}{Model} & \multicolumn{1}{l|}{Accuracy} & \multicolumn{1}{l|}{Precision} & \multicolumn{1}{l|}{Recall} & \multicolumn{1}{l|}{F1 Score} & \multicolumn{1}{l|}{ROC AUC} & \multicolumn{1}{l|}{Confusion Matrix} \\ \hline
\multicolumn{8}{|l|}{GPT2-Java} \\ \hline
sfd & GLM & 46.36\% & 0.4268 & 0.2121 & 0.2834 & 0.4622 & TN=118, FP=47, FN=130, TP=35 \\
ffd & GLM & 47.27\% & 0.4458 & 0.2242 & 0.2984 & 0.4787 & TN=119, FP=46, FN=128, TP=37 \\
gd & GLM & 46.06\% & 0.4316 & 0.2485 & 0.3154 & 0.4726 & TN=111, FP=54, FN=124, TP=41 \\
rpd & GLM & 62.42\% & 0.6395 & 0.5697 & 0.6026 & 0.6454 & TN=112, FP=53, FN=71, \hspace{4px}TP=94 \\ \hline
\multicolumn{8}{|l|}{CodeLLaMA} \\ \hline
sfd & GLM & 50.93\% & 0.5116 & 0.4074 & 0.4536 & 0.5103 & TN=99,\hspace{4px} FP=63, FN=96, TP=66 \\
ffd & GLM & 52.45\% & 0.527 & 0.4785 & 0.5016 & 0.4862 & TN=93,\hspace{4px} FP=70, FN=85, TP=78 \\
gd & GLM & 54.60\% & 0.5517 & 0.4908 & 0.5195 & 0.5211 & TN=98, \hspace{4px}FP=65, FN=83, TP=80 \\
rpd & GLM & 63.80\% & 0.7228 & 0.4479 & 0.553 & 0.6742 & TN=135, FP=28, FN=90, TP=73 \\ \hline
\end{tabular} \vspace{4px}
\label{tab:model-based-quartile-j}
\end{table*}

\begin{table*}[]
\centering
\vspace{-3mm}
\caption{Model-based analysis showing differentiability based on upper and lower quartiles ($V_{\text{H$\mu$}}^{q}$ vs. $V_{\text{L$\mu$}}^{q}$) for the Smith study (C).}
\vspace{-2mm}
\begin{tabular}{llllllll}
\hline
\multicolumn{1}{|l|}{Category} & \multicolumn{1}{l|}{Model} & \multicolumn{1}{l|}{Accuracy} & \multicolumn{1}{l|}{Precision} & \multicolumn{1}{l|}{Recall} & \multicolumn{1}{l|}{F1 Score} & \multicolumn{1}{l|}{ROC AUC} & \multicolumn{1}{l|}{Confusion Matrix} \\ \hline
\multicolumn{8}{|l|}{GPT2-C} \\ \hline
sfd & GLM & 57.61\% & 0.5686 & 0.6304 & 0.5979 & 0.5422 & TN=48, FP=44, FN=34, TP=58 \\
ffd & GLM & 56.45\% & 0.56 & 0.6022 & 0.5803 & 0.5356 & TN=49, FP=44, FN=37, TP=56 \\
gd & GLM & 56.45\% & 0.56 & 0.6022 & 0.5803 & 0.5389 & TN=49, FP=44, FN=37, TP=56 \\
rpd & GLM & 61.50\% & 0.604 & 0.6559 & 0.6289 & 0.6389 & TN=54, FP=40, FN=32, TP=61 \\ \hline
\multicolumn{8}{|l|}{CodeLLaMA} \\ \hline
sfd & GLM & 54.50\% & 0.5403 & 0.6032 & 0.57 & 0.5883 & TN=92,\hspace{4px} FP=97, FN=75, TP=114 \\
ffd & GLM & 52.34\% & 0.5224 & 0.5469 & 0.5344 & 0.5637 & TN=96,\hspace{4px} FP=96, FN=87, TP=105 \\
gd & GLM & 53.12\% & 0.5291 & 0.5677 & 0.5477 & 0.5756 & TN=95,\hspace{4px} FP=97, FN=83, TP=109 \\
rpd & GLM & 58.33\% & 0.5792 & 0.6094 & 0.5939 & 0.6127 & TN=107, FP=85, FN=75, TP=117 \\ \hline
\end{tabular} \vspace{4px}
\label{tab:model-based-quartile-c}
\vspace{-3mm}
\end{table*}

\vspace{-2mm}
\section{Model-based Results}
\label{sec:model-based-results}
\vspace{1mm}

Tables~\ref{tab:model-based-j} through~\ref{tab:model-based-quartile-c} summarize the results of our model-based analysis. Whereas the model-free analysis characterizes group-level differences in eye-tracking metrics, the model-based analysis evaluates whether SND and TF can predict whether a word belongs to a high- or low-attention group as defined by eye-tracking metrics.

\textbf{Median-based prediction of visual attention}
Tables~\ref{tab:model-based-j} and~\ref{tab:model-based-c} report results for prediction tasks in which words are labeled according to median splits on each eye-tracking metric. Across both the Java and C datasets, prediction performance based on SND and TF is modest for fixation-based measures, including single fixation duration (SFD), first fixation duration (FFD), and gaze duration (GD). Accuracies for these tasks generally fall near chance level (approximately 50--57\%), with ROC AUC values clustering around 0.50--0.55 for both GPT-2 and CodeLLaMA embeddings. 
Interestingly, this result suggests that regression path duration is more predictable from lexical properties at the individual-observation level, despite showing weaker group-level mean differences in the model-free analysis.  For the Java dataset, RPD prediction achieves accuracies exceeding 60\% and ROC AUC values approaching 0.67 under both embedding models. A similar, though weaker, trend is observed in the C dataset, where RPD consistently yields the highest predictive performance among the four eye-tracking metrics. 
Together, the model-free and model-based analyses indicate a dissociation between effect magnitude and predictability: early fixation measures exhibit clearer group-level differences, whereas regression path duration yields greater discriminability in predictive models.

\textbf{Upper- and lower-quartile prediction}
Tables~\ref{tab:model-based-quartile-j} and~\ref{tab:model-based-quartile-c} present results for prediction tasks using upper- and lower-quartile splits on each eye-tracking metric. By excluding observations near the median, this setting focuses on cases where high and low attention is more clear and less likely to be influenced by measurement noise. With quartile-based splits, predictive performance improves modestly across metrics, with the largest gains again observed for regression path duration. In the Java dataset, RPD prediction accuracy reaches approximately 63--64\%, with ROC AUC values near 0.67 for both embedding models. In the C dataset, RPD prediction similarly yields the strongest results, with ROC AUC values consistently above 0.60. Fixation-based metrics also show small improvements relative to median splits, but accuracies remain modest and AUC values generally remain near 0.55. These results reinforce those of the median-based prediction in that SND and TF exhibit limited but consistent predictive relationships with human visual attention, with clearer discriminability emerging when attention is defined using more extreme, high-contrast cases.


\textbf{Comparison across embedding models}
Across all prediction tasks, differences between GPT-2 and CodeLLaMA embeddings are relatively small. Both embedding models produce similar predictive trends, suggesting that the observed performance limits are not driven solely by embedding dimensionality, model scale, or even pretraining data. Instead, they likely reflect the inherently noisy nature of human visual attention during code reading.

\textbf{Summary.}
Overall, the model-based analysis shows that SND and TF yield limited but non-trivial predictive power for identifying words associated with higher versus lower levels of attention as measured by metrics of eye gaze time.  Predictive performance is consistently strongest for regression path duration, particularly under quartile-based splits, despite a lower observed difference of means in the model-free analysis.  Overall however, these findings complement the model-free analysis by demonstrating that SND and TF are associated with human visual attention.

\section{Discussion / Conclusion}
\label{sec:conclusion}

\begin{table*}[b]
\centering
\vspace{-3mm}
\caption{Key findings from other domains and consistency with our findings for Software Engineering tasks.}
\label{tab:findings}
\vspace{-2mm}
\begin{tabular}{llll}
                                  & Core Relevant Finding                                                & Domain                 &                \\
Buchanan \textit{et al}. (2001) \cite{Buchanan_2001a}   & Low SND + Low TF $\implies$ Slower Processing & Word Identification   & Not Consistent     \\
Mirman and Manguson (2008) \cite{Mirman_2008} & High Near SND $\implies$ Slower Processing               & Word Identification   & Consistent\\
Shaoul and Westbury (2010) \cite{Shaoul_2010a}   & High SND $\implies$ Slower Processing                 & Relatedness Decision   & Consistent \\
Al Farsi (2014) \cite{al2018survey}     & High SND + High TF $\implies$ Faster Processing                & Reading Sentences   & Neutral \\
Fieder \textit{et al}. (2019) \cite{fieder_close_2019}    & High SND $\implies$ Slower Processing                 & Picture Identification   & Neutral     \\
\end{tabular}
\end{table*}

The purpose of this paper is to determine if there is an effect of SND or TF on human visual attention during software engineering tasks.  In our model-free analysis we interpret FDR-corrected statistically significant differences as signs of a meaningful effect.  Likewise in our model-based analysis, we view \textbf{any} degree, however slight, of predictive power as a positive result.  We do \textit{not} claim that the model we trained is a highly effective or complete predictor of human attention.  In fact, we would view an excessively strong result with suspicion; our model only has two inputs, and eye gaze metrics are notoriously noisy and subject to the personal background and vicissitudes of life of the participants.  Still, our finding that relatively easy-to-calculate lexical factors are related to human visual attention has implications for software engineering research as well as academic value in understanding human programmer cognition.


\textbf{Scope Summary} We emphasize the scope of this paper in the software engineering domain: we study SND over two programming languages and experimental data involving different programming tasks.  We find that one common way to measure SND (ARC, see Section~\ref{sec:snd}) is associated with human attention in this domain.  However, there are many other ways to calculate SND that are beyond the scope of this one paper.  We caution that the discussion of our research contributions below should be viewed with this scope in mind and may not apply to all forms of SND.

\textbf{Software Engineering Perspective} From a software engineering perspective, this work is useful for two reasons.  First, we provide new knowledge about how programmers read source code and the factors influencing their visual attention.  Beyond the academic value of understanding human behavior, this knowledge could help designers of user interfaces to present information in a way that is simpler to read~\cite{armaly2018audiohighlight, hejmady2012visual}, help guide authors of programming style guides~\cite{de2024assessing}, and help educators understand where people may need to spend more attentional effort~\cite{blackwell2002first}.  Second, our findings could help designers of models of programmer attention by describing the effect of possible feature inputs to those models.  These models in turn help make AI models for software engineering tasks act more human-like~\cite{bansal2023human}.

\textbf{Cognitive Theory Perspective} From a cognitive theory perspective, this work contributes to an ongoing discussion about the relationship of SND to human attention in various domains.  Words in these domains are understood to be related to each other in different ways, leading to different methods of computing SND~\cite{AlFarsi2014}.  These differences lead to conclusions to which our findings are consistent, not consistent, and neutral.  We summarize some of these conclusions in Table~\ref{tab:findings} and describe how they relate to our work below.


Buchanan \textit{et al.}~\cite{Buchanan_2001a} computed SND as the mean distance between a word and the ten nearest semantic neighbors, using reaction time rather than fixation duration to capture human reading time. When presenting the participant with a word and asking for a rapid identification of that word, they found a small semantic neighborhood (high neighborhood density) facilitated reading time for low frequency words, while they observed no correlation between high frequency words and SND. Al Farsi~\cite{AlFarsi2014} studied sentence reading and comprehension, and found that high SND decreased reading times, but found that high frequency words were facilitatory. Low SND correlated with inhibited reading time, especially for high-frequency words, though primarily in late reading time measures, such as regression path duration. Our findings in the C dataset confirm that high-frequency words receive shorter fixation durations, but find that SND rather has an opposite effect in a programming context, with high SND correlating with longer fixation duration on the C dataset, and on both datasets when combined with low frequency. 

Shaoul and Westbury~\cite{Shaoul_2010a} introduced ARC as a new measurement for SND, and also found that words with higher SND had slower reading times. This study used a ``relatedness decision task'', which in this case just means comparing two words. Mirman and Magnuson~\cite{Mirman_2008} found a similar result when looking at dense near neighborhoods and faster for dense distant neighborhoods, using a task that either asked the participant to identify real words or determine if a thing was living or non-living. Fieder \textit{et al.}~\cite{fieder_close_2019} found that in image recognition, a high density of near semantic neighbors increased latency and decreased accuracy. Our findings confirm these results in a programming context, finding high SND to be inhibitory in C, and high SND and frequency to be inhibitory in both C and Java. 

Several arguments for the differences in results across the discipline exist, which could have implications for our future work. Mirman and Magnuson's results could imply that our measure of SND captures more near neighbors, inhibiting processing speed, while papers such as Reilly and Desai~\cite{REILLY201746} suggest to use that emotional arousal of abstract words could play a role in impeding processing speed.  We propose for future work that a deeper analysis of the semantic space of code could improve our understanding of the relationship between SND and human eye gaze time.

\begin{figure}[t]
\vspace{4px}
\centering
{
    \includegraphics[width=9cm]{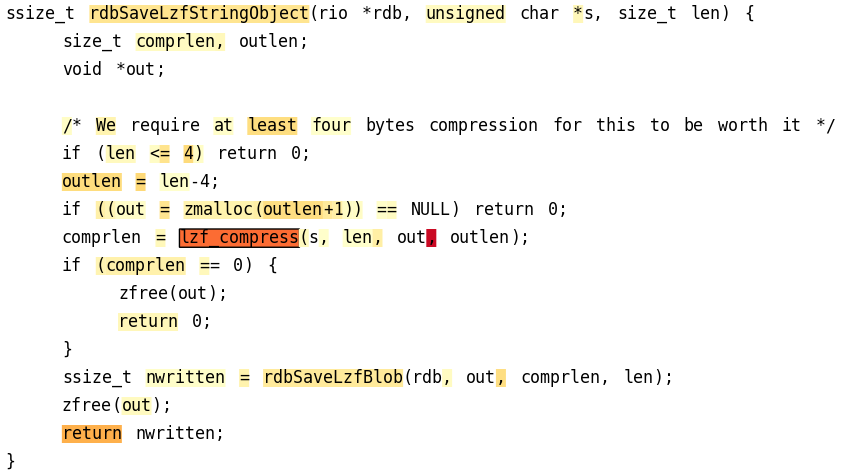} \\
}
\vspace{-5px}
\caption{Example of fixations on a C dataset task. Darker highlights indicate longer fixation duration. Boxed tokens fall in the $V_{\text{HSND,LF}}$ category}
\label{fig:fixation_example}
\vspace{-5mm}
\end{figure}

\textbf{Example} Figure \ref{fig:fixation_example} shows the fixations of participant 10 on the {\tt{stonefly}} task from the C study, while trying to locate an integer overflow. The highlighted tokens are ones which have a fixation duration, and the darker the highlight, the longer the measured fixation. Tokens with a box drawn around them are in the high SND, low frequency category. From the information the participant is given, they are tasked with finding the line which contains a bug, and know that the line is either in this function, or in a function called here. Of the functions called, {\tt{lzf\_compress}} and {\tt{rdbSaveLzfBlob}} are the most likely candidates, as the other functions deal with allocating memory, which is a different type of error. Additionally, the participant is informed that the bug has to do with a returned value, so we see longer attention near the return value of this function. In the context of the assigned bug-finding task, the token {\tt{lzf\_compress}} is only important for comprehending what the function {\tt{rdbSaveLzfStringObject}} does. But the bug itself is in {\tt{rdbWriteRaw}}, which is called by {\tt{rdbSaveLzfBlob}} in this example. So, {\tt{lzf\_compress}} is an example of a token that serves no direct purpose for the task, which could possibly be an example of high SND \& low frequency having an effect on reading time.

\textbf{Limitations}
As with any study, this work carries limitations and threats to validity. First, while eye-tracking metrics are widely used as proxies for human visual attention, but they do not directly measure comprehension or cognitive state. Second, we derived SND from distributional word embeddings and these could reflect statistical regularities in a corpus rather than ground-truth semantics.  Alternative embeddings or SND formulations could yield different values. Third, although we analyze two programming languages, two tasks, and multiple embedding models, our datasets are limited to specific repositories and participant populations, which may affect generalizability. Finally, our word-level analysis abstracts away higher-level code structure and context as well as other visual cues that could have been present during the Smith and Wallace studies.  Data from different studies may lead to different conclusions.

\textbf{Reproducibility}
To support ongoing work and reproducibility, we release all data, scripts, and results via an online appendix:

\vspace{2mm}
\url{https://github.com/apcl-research/attention-semantics}


\section{Acknowledgments}
This work is supported in part by the NSF grants NSF CCF-2100035 and CCF-2211428. 
Any opinions, findings, and conclusions expressed herein are the authors’ and do not necessarily reflect those of the sponsors.
\vspace{2mm}

\bibliographystyle{IEEEtran}
\bibliography{IEEEabrv,main}

\begin{thebibliography}{10}
\providecommand{\url}[1]{#1}
\csname url@samestyle\endcsname
\providecommand{\newblock}{\relax}
\providecommand{\bibinfo}[2]{#2}
\providecommand{\BIBentrySTDinterwordspacing}{\spaceskip=0pt\relax}
\providecommand{\BIBentryALTinterwordstretchfactor}{4}
\providecommand{\BIBentryALTinterwordspacing}{\spaceskip=\fontdimen2\font plus
\BIBentryALTinterwordstretchfactor\fontdimen3\font minus \fontdimen4\font\relax}
\providecommand{\BIBforeignlanguage}[2]{{%
\expandafter\ifx\csname l@#1\endcsname\relax
\typeout{** WARNING: IEEEtran.bst: No hyphenation pattern has been}%
\typeout{** loaded for the language `#1'. Using the pattern for}%
\typeout{** the default language instead.}%
\else
\language=\csname l@#1\endcsname
\fi
#2}}
\providecommand{\BIBdecl}{\relax}
\BIBdecl

\bibitem{mole2025encylopedia}
C.~Mole, ``{Attention},'' in \emph{The {Stanford} Encyclopedia of Philosophy}, E.~N. Zalta and U.~Nodelman, Eds.\hskip 1em plus 0.5em minus 0.4em\relax Metaphysics Research Lab, Stanford University, 2025.

\bibitem{novak2024eye}
J.~{\v{S}}. Nov{\'a}k, J.~Masner, P.~Benda, P.~{\v{S}}imek, and V.~Merunka, ``Eye tracking, usability, and user experience: A systematic review,'' \emph{International Journal of Human--Computer Interaction}, vol.~40, no.~17, pp. 4484--4500, 2024.

\bibitem{baharum2024enhancing}
A.~Baharum, R.~Ismail, S.~Halamy, E.~A. Rahim, N.~A.~M. Noor, and F.~D. Deris, ``Enhancing ux design through eye-tracking and image processing: Practical insights and applications,'' in \emph{2024 International Conference on Platform Technology and Service (PlatCon)}.\hskip 1em plus 0.5em minus 0.4em\relax IEEE, 2024, pp. 1--4.

\bibitem{cheng2012eye}
S.~Cheng and Y.~Liu, ``Eye-tracking based adaptive user interface: implicit human-computer interaction for preference indication,'' \emph{Journal on Multimodal User Interfaces}, vol.~5, pp. 77--84, 2012.

\bibitem{zhang2024eyetrans}
Y.~Zhang, J.~Li, Z.~Karas, A.~Bansal, T.~J.-J. Li, C.~McMillan, K.~Leach, and Y.~Huang, ``Eyetrans: Merging human and machine attention for neural code summarization,'' in \emph{Proceedings of The ACM Joint European Software Engineering Conference and Symposium on the Foundations of Software Engineering (ESEC/FSE 2024)}, 2024.

\bibitem{zhang2025enhancing}
Y.~Zhang, C.~Huang, Z.~Karas, D.~T. Nguyen, K.~Leach, and Y.~Huang, ``Enhancing code llm training with programmer attention,'' \emph{arXiv preprint arXiv:2503.14936}, 2025.

\bibitem{pourhosein2025unveiling}
M.~Pourhosein and M.~Sabokro, ``Unveiling the gaze: deciphering key factors in selecting knowledge workers through eye-tracking analysis,'' \emph{European Journal of Management Studies}, 2025.

\bibitem{kiseleva2020study}
E.~Kiseleva, E.~Gudoshnik, A.~Orlov, and A.~Rustemova, ``Study of the possibility of using pupillography for personnel selection at hiring,'' in \emph{Journal of physics: conference series}, vol. 1519, no.~1.\hskip 1em plus 0.5em minus 0.4em\relax IOP Publishing, 2020, p. 012023.

\bibitem{Buchanan_2001a}
\BIBentryALTinterwordspacing
L.~Buchanan, C.~Westbury, and C.~Burgess, ``Characterizing semantic space: {{Neighborhood}} effects in word recognition,'' \emph{Psychonomic Bulletin \& Review}, vol.~8, no.~3, pp. 531--544, 2001. [Online]. Available: \url{http://link.springer.com/10.3758/BF03196189}
\BIBentrySTDinterwordspacing

\bibitem{Mirman_2008}
D.~Mirman and J.~S. Magnuson, ``Attractor dynamics and semantic neighborhood density: {{Processing}} is slowed by near neighbors and speeded by distant neighbors.'' \emph{Journal of Experimental Psychology: Learning, Memory, and Cognition}, vol.~34, no.~1, pp. 65--79, 2008.

\bibitem{maalej2014comprehension}
W.~Maalej, R.~Tiarks, T.~Roehm, and R.~Koschke, ``On the comprehension of program comprehension,'' \emph{ACM Transactions on Software Engineering and Methodology (TOSEM)}, vol.~23, no.~4, pp. 1--37, 2014.

\bibitem{schroter2017comprehending}
I.~Schr{\"o}ter, J.~Kr{\"u}ger, J.~Siegmund, and T.~Leich, ``Comprehending studies on program comprehension,'' in \emph{2017 IEEE/ACM 25th International Conference on Program Comprehension (ICPC)}.\hskip 1em plus 0.5em minus 0.4em\relax IEEE, 2017, pp. 308--311.

\bibitem{Danguecan_2016}
A.~N. Danguecan and L.~Buchanan, ``Semantic {{Neighborhood Effects}} for {{Abstract}} versus {{Concrete Words}},'' \emph{Frontiers in Psychology}, vol.~7, Jul. 2016.

\bibitem{REILLY201746}
\BIBentryALTinterwordspacing
M.~Reilly and R.~H. Desai, ``Effects of semantic neighborhood density in abstract and concrete words,'' \emph{Cognition}, vol. 169, pp. 46--53, 2017. [Online]. Available: \url{https://www.sciencedirect.com/science/article/pii/S0010027717302226}
\BIBentrySTDinterwordspacing

\bibitem{Burgess1998}
C.~Burgess, ``From simple associations to the building blocks of language: {{Modeling}} meaning in memory with the {{HAL}} model,'' \emph{Behavior Research Methods, Instruments, \& Computers}, vol.~30, no.~2, pp. 188--198, Jun. 1998.

\bibitem{Shaoul_2010a}
C.~Shaoul and C.~Westbury, ``Exploring lexical co-occurrence space using {{HiDEx}},'' \emph{Behavior Research Methods}, vol.~42, no.~2, pp. 393--413, May 2010.

\bibitem{AlFarsi2014}
B.~Al~Farsi, ``Semantic neighbourhood density effects in word identification during normal reading: evidence from eye movements,'' Ph.D. dissertation, University of Southampton, 08 2014.

\bibitem{rodeghero2014improving}
P.~Rodeghero, C.~McMillan, P.~W. McBurney, N.~Bosch, and S.~D'Mello, ``Improving automated source code summarization via an eye-tracking study of programmers,'' in \emph{Proceedings of the 36th international conference on Software engineering}.\hskip 1em plus 0.5em minus 0.4em\relax ACM, 2014, pp. 390--401.

\bibitem{Pennington2014GloVeGV}
\BIBentryALTinterwordspacing
J.~Pennington, R.~Socher, and C.~D. Manning, ``Glove: Global vectors for word representation,'' in \emph{Conference on Empirical Methods in Natural Language Processing}, 2014. [Online]. Available: \url{https://api.semanticscholar.org/CorpusID:1957433}
\BIBentrySTDinterwordspacing

\bibitem{al2021novice}
N.~Al~Madi, C.~S. Peterson, B.~Sharif, and J.~I. Maletic, ``From novice to expert: Analysis of token level effects in a longitudinal eye tracking study,'' in \emph{2021 IEEE/ACM 29th International Conference on Program Comprehension (ICPC)}.\hskip 1em plus 0.5em minus 0.4em\relax IEEE, 2021, pp. 172--183.

\bibitem{abid2019developer}
N.~J. Abid, B.~Sharif, N.~Dragan, H.~Alrasheed, and J.~I. Maletic, ``Developer reading behavior while summarizing java methods: Size and context matters,'' in \emph{2019 IEEE/ACM 41st International Conference on Software Engineering (ICSE)}.\hskip 1em plus 0.5em minus 0.4em\relax IEEE, 2019, pp. 384--395.

\bibitem{Sharafi2020PracticalGuide}
\BIBentryALTinterwordspacing
Z.~Sharafi, B.~Sharif, Y.~Gu{\'{e}}h{\'{e}}neuc, A.~Begel, R.~Bednarik, and M.~E. Crosby, ``A practical guide on conducting eye tracking studies in software engineering,'' \emph{Empir. Softw. Eng.}, vol.~25, no.~5, pp. 3128--3174, 2020. [Online]. Available: \url{https://doi.org/10.1007/s10664-020-09829-4}
\BIBentrySTDinterwordspacing

\bibitem{Wallace2025Programmer}
R.~Wallace, A.~Bansal, Z.~Karas, N.~Tang, Y.~Huang, T.~J.-J. Li, and C.~McMillan, ``Programmer visual attention during context-aware code summarization,'' \emph{IEEE Transactions on Software Engineering}, pp. 1--13, 2025.

\bibitem{Smith_2025}
\BIBentryALTinterwordspacing
E.~Smith, R.~Wallace, M.~Robison, Y.~Huang, and C.~McMillan, ``Human attention during localization of memory bugs in c programs,'' preprint. [Online]. Available: \url{https://arxiv.org/abs/2506.00693}
\BIBentrySTDinterwordspacing

\bibitem{grabinger2024eye}
L.~Grabinger, F.~Hauser, C.~Wolff, and J.~Mottok, ``On eye tracking in software engineering,'' \emph{SN Computer Science}, vol.~5, no.~6, p. 729, 2024.

\bibitem{Obaidellah2018}
\BIBentryALTinterwordspacing
U.~Obaidellah, M.~A. Haek, and P.~C. Cheng, ``A survey on the usage of eye-tracking in computer programming,'' \emph{{ACM} Comput. Surv.}, vol.~51, no.~1, pp. 5:1--5:58, 2018. [Online]. Available: \url{https://doi.org/10.1145/3145904}
\BIBentrySTDinterwordspacing

\bibitem{sharif2011use}
B.~Sharif and H.~Kagdi, ``On the use of eye tracking in software traceability,'' in \emph{Proceedings of the 6th International Workshop on Traceability in Emerging Forms of Software Engineering}, 2011, pp. 67--70.

\bibitem{Sharif2017Traceability}
\BIBentryALTinterwordspacing
B.~Sharif, J.~Meinken, T.~Shaffer, and H.~H. Kagdi, ``Eye movements in software traceability link recovery,'' \emph{Empir. Softw. Eng.}, vol.~22, no.~3, pp. 1063--1102, 2017. [Online]. Available: \url{https://doi.org/10.1007/s10664-016-9486-9}
\BIBentrySTDinterwordspacing

\bibitem{behler2023itracetoolkit}
J.~Behler, P.~Weston, D.~T. Guarnera, B.~Sharif, and J.~I. Maletic, ``itrace-toolkit: A pipeline for analyzing eye-tracking data of software engineering studies,'' in \emph{2023 IEEE/ACM 45th International Conference on Software Engineering: Companion Proceedings (ICSE-Companion)}, 2023, pp. 46--50.

\bibitem{winter2024iconicity}
B.~Winter, G.~Lupyan, L.~K. Perry, M.~Dingemanse, and M.~Perlman, ``Iconicity ratings for 14,000+ english words,'' \emph{Behavior research methods}, vol.~56, no.~3, pp. 1640--1655, 2024.

\bibitem{giraldez2020effect}
M.~G. Elizo, ``The effect of semantic neighborhood density on vocabulary learning in spanish as a second language and spanish as a heritage language,'' Ph.D. dissertation, The University of New Mexico, 2020.

\bibitem{ayasse2020two}
N.~D. Ayasse and A.~Wingfield, ``The two sides of linguistic context: Eye-tracking as a measure of semantic competition in spoken word recognition among younger and older adults,'' \emph{Frontiers in Human Neuroscience}, vol.~14, p. 132, 2020.

\bibitem{harel2021age}
T.~Harel-Arbeli, A.~Wingfield, Y.~Palgi, and B.~M. Ben-David, ``Age-related differences in the online processing of spoken semantic context and the effect of semantic competition: evidence from eye gaze,'' \emph{Journal of Speech, Language, and Hearing Research}, vol.~64, no.~2, pp. 315--327, 2021.

\bibitem{tamminen2013role}
J.~Tamminen, M.~A.~L. Ralph, and P.~A. Lewis, ``The role of sleep spindles and slow-wave activity in integrating new information in semantic memory,'' \emph{Journal of Neuroscience}, vol.~33, no.~39, pp. 15\,376--15\,381, 2013.

\bibitem{sun2014empirical}
X.~Sun, X.~Liu, J.~Hu, and J.~Zhu, ``Empirical studies on the nlp techniques for source code data preprocessing,'' in \emph{Proceedings of the 2014 3rd international workshop on evidential assessment of software technologies}, 2014, pp. 32--39.

\bibitem{collard2011lightweight}
M.~L. Collard, M.~J. Decker, and J.~I. Maletic, ``Lightweight transformation and fact extraction with the srcml toolkit,'' in \emph{2011 IEEE 11th international working conference on source code analysis and manipulation}.\hskip 1em plus 0.5em minus 0.4em\relax IEEE, 2011, pp. 173--184.

\bibitem{butler2016analysing}
S.~Butler, \emph{Analysing Java Identifier Names}.\hskip 1em plus 0.5em minus 0.4em\relax Open University (United Kingdom), 2016.

\bibitem{herka2023identifier}
I.~Herka, ``Identifier names in computer programs: Literature review.'' \emph{Advances in Cognitive Psychology}, vol.~19, no.~3, 2023.

\bibitem{scanniello2017fixing}
G.~Scanniello, M.~Risi, P.~Tramontana, and S.~Romano, ``Fixing faults in c and java source code: Abbreviated vs. full-word identifier names,'' \emph{ACM Transactions on Software Engineering and Methodology (TOSEM)}, vol.~26, no.~2, pp. 1--43, 2017.

\bibitem{su2023language}
C.-Y. Su, A.~Bansal, V.~Jain, S.~Ghanavati, and C.~Mcmillan, ``A language model of java methods with train/test deduplication,'' \emph{arXiv preprint arXiv:2305.08286}, 2023.

\bibitem{su2024distilled}
C.-Y. Su and C.~McMillan, ``Distilled gpt for source code summarization,'' \emph{Automated Software Engineering}, vol.~31, no.~1, p.~22, 2024.

\bibitem{roziere2023code}
B.~Roziere, J.~Gehring, F.~Gloeckle, S.~Sootla, I.~Gat, X.~E. Tan, Y.~Adi, J.~Liu, R.~Sauvestre, T.~Remez \emph{et~al.}, ``Code llama: Open foundation models for code,'' \emph{arXiv preprint arXiv:2308.12950}, 2023.

\bibitem{clifton2007eye}
C.~Clifton~Jr, A.~Staub, and K.~Rayner, ``Eye movements in reading words and sentences,'' \emph{Eye movements}, pp. 341--371, 2007.

\bibitem{rayner1998eye}
K.~Rayner, ``Eye movements in reading and information processing: 20 years of research.'' \emph{Psychological bulletin}, vol. 124, no.~3, p. 372, 1998.

\bibitem{rayner2004effect}
K.~Rayner, T.~Warren, B.~J. Juhasz, and S.~P. Liversedge, ``The effect of plausibility on eye movements in reading.'' \emph{Journal of Experimental Psychology: Learning, Memory, and Cognition}, vol.~30, no.~6, p. 1290, 2004.

\bibitem{warren2007investigating}
T.~Warren and K.~McConnell, ``Investigating effects of selectional restriction violations and plausibility violation severity on eye-movements in reading,'' \emph{Psychonomic bulletin \& review}, vol.~14, no.~4, pp. 770--775, 2007.

\bibitem{staub2010eye}
A.~Staub, ``Eye movements and processing difficulty in object relative clauses,'' \emph{Cognition}, vol. 116, no.~1, pp. 71--86, 2010.

\bibitem{good2005permutation}
P.~Good, \emph{Permutation, parametric and bootstrap tests of hypotheses}.\hskip 1em plus 0.5em minus 0.4em\relax Springer, 2005.

\bibitem{ernst2004permutation}
M.~D. Ernst, ``Permutation methods: a basis for exact inference,'' \emph{Statistical Science}, pp. 676--685, 2004.

\bibitem{krajewski2010rh}
G.~Krajewski and D.~Matthews, ``Rh baayen, analyzing linguistic data: A practical introduction to statistics using r. cambridge: Cambridge university press, 2008. pp. 368. isbn-13: 978-0-521-70918-7.'' \emph{Journal of Child Language}, vol.~37, no.~2, pp. 465--470, 2010.

\bibitem{jaeger2008categorical}
T.~F. Jaeger, ``Categorical data analysis: Away from anovas (transformation or not) and towards logit mixed models,'' \emph{Journal of Memory and Language}, vol.~59, no.~4, pp. 434--446, 2008.

\bibitem{fox2015applied}
J.~Fox, \emph{Applied regression analysis and generalized linear models}.\hskip 1em plus 0.5em minus 0.4em\relax Sage publications, 2015.

\bibitem{arlot2010survey}
S.~Arlot and A.~Celisse, ``A survey of cross-validation procedures for model selection,'' \emph{Statistics Surveys}, vol.~4, pp. 40--79, 2010.

\bibitem{kliegl2004length}
R.~Kliegl, E.~Grabner, M.~Rolfs, and R.~Engbert, ``Length, frequency, and predictability effects of words on eye movements in reading,'' \emph{European journal of cognitive psychology}, vol.~16, no. 1-2, pp. 262--284, 2004.

\bibitem{starke2009searching}
J.~Starke, C.~Luce, and J.~Sillito, ``Searching and skimming: An exploratory study,'' in \emph{2009 IEEE International Conference on Software Maintenance}.\hskip 1em plus 0.5em minus 0.4em\relax IEEE, 2009, pp. 157--166.

\bibitem{al2018survey}
A.~Al-Kaff, D.~Martin, F.~Garcia, A.~de~la Escalera, and J.~M. Armingol, ``Survey of computer vision algorithms and applications for unmanned aerial vehicles,'' \emph{Expert Systems with Applications}, vol.~92, pp. 447--463, 2018.

\bibitem{fieder_close_2019}
\BIBentryALTinterwordspacing
N.~Fieder, I.~Wartenburger, and R.~Abdel~Rahman, ``A close call: {Interference} from semantic neighbourhood density and similarity in language production,'' \emph{Memory \& Cognition}, vol.~47, no.~1, pp. 145--168, Jan. 2019. [Online]. Available: \url{https://doi.org/10.3758/s13421-018-0856-y}
\BIBentrySTDinterwordspacing

\bibitem{armaly2018audiohighlight}
A.~Armaly, P.~Rodeghero, and C.~McMillan, ``Audiohighlight: Code skimming for blind programmers,'' in \emph{2018 IEEE International Conference on Software Maintenance and Evolution (ICSME)}.\hskip 1em plus 0.5em minus 0.4em\relax IEEE, 2018, pp. 206--216.

\bibitem{hejmady2012visual}
P.~Hejmady and N.~H. Narayanan, ``Visual attention patterns during program debugging with an ide,'' in \emph{proceedings of the symposium on eye tracking research and applications}, 2012, pp. 197--200.

\bibitem{de2024assessing}
P.~R.~F. de~Oliveira, R.~Gheyi, J.~A.~S. da~Costa, and M.~Ribeiro, ``Assessing python style guides: An eye-tracking study with novice developers,'' in \emph{Simp{\'o}sio Brasileiro de Engenharia de Software (SBES)}.\hskip 1em plus 0.5em minus 0.4em\relax SBC, 2024, pp. 136--146.

\bibitem{blackwell2002first}
A.~F. Blackwell, ``First steps in programming: A rationale for attention investment models,'' in \emph{Proceedings IEEE 2002 Symposia on Human Centric Computing Languages and Environments}.\hskip 1em plus 0.5em minus 0.4em\relax IEEE, 2002, pp. 2--10.

\bibitem{bansal2023human}
A.~Bansal, B.~Sharif, and C.~McMillan, ``Towards modeling human attention from eye movements for neutral source code summarization,'' \emph{Proceedings of ACM Human-Computer Interaction, ETRA Vol. 7}, 2023.

\end{thebibliography}

\end{document}